\documentclass[11pt]{article}
\usepackage{epsfig}
\usepackage{rotating}
\usepackage{amsmath}
\usepackage{hhline}
\usepackage{amssymb}
\newlength{\dinwidth}
\newlength{\dinmargin}
\begin{document}  
\newcommand{\pom}{{I\!\!P}}
\newcommand{\slowpi}{\pi_{\mathit{slow}}}
\newcommand{\fiidiii}{F_2^{D(3)}}
\newcommand{\fiidiiiarg}{\fiidiii\,(\beta,\,Q^2,\,x)}
\newcommand{\n}{1.19\pm 0.06 (stat.) \pm0.07 (syst.)}
\newcommand{\nz}{1.30\pm 0.08 (stat.)^{+0.08}_{-0.14} (syst.)}
\newcommand{\fiidiiiful}{F_2^{D(4)}\,(\beta,\,Q^2,\,x,\,t)}
\newcommand{\fiipom}{\tilde F_2^D}
\newcommand{\ALPHA}{1.10\pm0.03 (stat.) \pm0.04 (syst.)}
\newcommand{\ALPHAZ}{1.15\pm0.04 (stat.)^{+0.04}_{-0.07} (syst.)}
\newcommand{\fiipomarg}{\fiipom\,(\beta,\,Q^2)}
\newcommand{\pomflux}{f_{\pom / p}}
\newcommand{\nxpom}{1.19\pm 0.06 (stat.) \pm0.07 (syst.)}
\newcommand {\gapprox}
   {\raisebox{-0.7ex}{$\stackrel {\textstyle>}{\sim}$}}
\newcommand {\lapprox}
   {\raisebox{-0.7ex}{$\stackrel {\textstyle<}{\sim}$}}
\def\gsim{\,\lower.25ex\hbox{$\scriptstyle\sim$}\kern-1.30ex%
\raise 0.55ex\hbox{$\scriptstyle >$}\,}
\def\lsim{\,\lower.25ex\hbox{$\scriptstyle\sim$}\kern-1.30ex%
\raise 0.55ex\hbox{$\scriptstyle <$}\,}
\newcommand{\pomfluxarg}{f_{\pom / p}\,(x_\pom)}
\newcommand{\dsf}{\mbox{$F_2^{D(3)}$}}
\newcommand{\dsfva}{\mbox{$F_2^{D(3)}(\beta,Q^2,x_{I\!\!P})$}}
\newcommand{\dsfvb}{\mbox{$F_2^{D(3)}(\beta,Q^2,x)$}}
\newcommand{\dsfpom}{$F_2^{I\!\!P}$}
\newcommand{\gap}{\stackrel{>}{\sim}}
\newcommand{\lap}{\stackrel{<}{\sim}}
\newcommand{\fem}{$F_2^{em}$}
\newcommand{\tsnmp}{$\tilde{\sigma}_{NC}(e^{\mp})$}
\newcommand{\tsnm}{$\tilde{\sigma}_{NC}(e^-)$}
\newcommand{\tsnp}{$\tilde{\sigma}_{NC}(e^+)$}
\newcommand{\st}{$\star$}
\newcommand{\sst}{$\star \star$}
\newcommand{\ssst}{$\star \star \star$}
\newcommand{\sssst}{$\star \star \star \star$}
\newcommand{\tw}{\theta_W}
\newcommand{\sw}{\sin{\theta_W}}
\newcommand{\cw}{\cos{\theta_W}}
\newcommand{\sww}{\sin^2{\theta_W}}
\newcommand{\cww}{\cos^2{\theta_W}}
\newcommand{\trm}{m_{\perp}}
\newcommand{\trp}{p_{\perp}}
\newcommand{\trmm}{m_{\perp}^2}
\newcommand{\trpp}{p_{\perp}^2}
\newcommand{\alp}{\alpha_s}

\newcommand{\alps}{\alpha_s}
\newcommand{\sqrts}{$\sqrt{s}$}
\newcommand{\LO}{$O(\alpha_s^0)$}
\newcommand{\Oa}{$O(\alpha_s)$}
\newcommand{\Oaa}{$O(\alpha_s^2)$}
\newcommand{\PT}{p_{\perp}}
\newcommand{\JPSI}{J/\psi}
\newcommand{\sh}{\hat{s}}
\newcommand{\uh}{\hat{u}}
\newcommand{\MP}{m_{J/\psi}}
\newcommand{\PO}{I\!\!P}
\newcommand{\xbj}{x}
\newcommand{\xpom}{x_{\PO}}
\newcommand{\ttbs}{\char'134}
\newcommand{\xpomlo}{3\times10^{-4}}  
\newcommand{\xpomup}{0.05}  
%
%
\newcommand{\qsq}{\ensuremath{Q^2} }
\newcommand{\xg}{x_{\gamma} }
\newcommand{\xgr}{x_{\gamma}^{rec} }
\newcommand{\gevsq}{\ensuremath{\mathrm{GeV}^2} }
\newcommand{\et}{\ensuremath{E_t^*} }
\newcommand{\rap}{\ensuremath{\eta^*} }
\newcommand{\gp}{\ensuremath{\gamma^*}p }
\newcommand{\dsiget}{\ensuremath{{\rm d}\sigma_{ep}/{\rm d}E_t^*} }
\newcommand{\dsigrap}{\ensuremath{{\rm d}\sigma_{ep}/{\rm d}\eta^*} }

\begin{titlepage}
\begin{flushleft}
{\tt DESY 98-148    \hfill    ISSN 0418-9833} \\
{\tt October 1998}                  \\
\end{flushleft}
\vspace*{3.0cm}
\begin{center}\begin{LARGE}
{\bf Charged Particle Cross Sections in Photoproduction and Extraction of
the 
Gluon Density in the Photon} \\

\vspace*{2.5cm}
H1 Collaboration \\
\vspace*{2.5cm}
\end{LARGE}
{\bf Abstract}
\begin{quotation}
\noindent
Photoproduction data collected with the H1 detector at HERA in 1994 are used 
to study the cross-sections for inclusive charged particle production and 
the structure of the photon.  The differential cross-sections 
${\rm d}\sigma/{\rm d}p_T^2, $
for $|\eta| < 1$ in the HERA laboratory frame, and 
${\rm d}\sigma/{\rm d}\eta$ for 
$p_T> 2$ GeV/c and  $p_T> 3$ GeV/c
have been measured. 
Model calculations of
these cross-sections, based on perturbative QCD, indicate that the results
are sensitive to the parton densities of the photon as well as to higher
order effects, which are phenomenologically treated by multiple
interactions. This sensitivity is exploited to determine the leading order
$x_{\gamma}$ distribution of partons in the photon
using a new method based on  high $p_T$ charged
particles. 
The 
gluon content of the photon is extracted and
  found to rise with decreasing $x_{\gamma}$.

\end{quotation}
\vspace*{2.0cm}
 Submitted to Eur. Phys. J. 
\end{center}
\cleardoublepage
\end{titlepage}

\begin{Large} \begin{center} H1 Collaboration \end{center} \end{Large}
\begin{flushleft}
 C.~Adloff$^{34}$,                
 M.~Anderson$^{22}$,              
 V.~Andreev$^{25}$,               
 B.~Andrieu$^{28}$,               
 V.~Arkadov$^{35}$,               
 C.~Arndt$^{11}$,                 
 I.~Ayyaz$^{29}$,                 
 A.~Babaev$^{24}$,                
 J.~B\"ahr$^{35}$,                
 J.~B\'an$^{17}$
% ...

\end{flushleft}
\begin{flushleft} {\it
% ...
$ ^2$ III. Physikalisches Institut der RWTH, Aachen, Germany$^a$ \\
% ...
$ ^7$ Physics Department and IIRPA,
       University of California, Davis, California, USA$^e$ \\
% ...
$ ^{11}$ DESY, Hamburg, Germany$^a$ \\
% ...
$ ^{17}$ Institute of Experimental Physics, Slovak Academy of
          Sciences, Ko\v{s}ice, Slovak Republic$^{f,j}$ \\
% ...
$ ^{22}$ Department of Physics and Astronomy, 
          University of Manchester, Manchester, UK$^b$ \\
% ...
$ ^{24}$ Institute for Theoretical and Experimental Physics,
          Moscow, Russia \\
 $ ^{25}$ Lebedev Physical Institute, Moscow, Russia$^{f,k}$ \\
% ...
$ ^{28}$ LPNHE, \'{E}cole Polytechnique, IN2P3-CNRS, Palaiseau, France \\
 $ ^{29}$ LPNHE, Universit\'{e}s Paris VI and VII, IN2P3-CNRS,
          Paris, France \\
% ...
$ ^{34}$ Fachbereich Physik, Bergische Universit\"at Gesamthochschule
          Wuppertal, Wuppertal, Germany$^a$ \\
 $ ^{35}$ DESY, Institut f\"ur Hochenergiephysik, Zeuthen, Germany$^a$ \\
% ...
$ ^a$ Supported by the Bundesministerium f\"ur Bildung, Wissenschaft,
        Forschung und Technologie, FRG,
        under contract numbers 7AC17P, 7AC47P, 7DO55P, 7HH17I, 7HH27P,
        7HD17P, 7HD27P, 7KI17I, 6MP17I and 7WT87P \\
 $ ^b$ Supported by the UK Particle Physics and Astronomy Research
       Council, and formerly by the UK Science and Engineering Research
       Council \\
% ...
$ ^e$ Supported in part by US~DOE grant DE~F603~91ER40674 \\
 $ ^f$ Supported by the Deutsche Forschungsgemeinschaft \\
% ...
$ ^j$ Supported by VEGA SR grant no. 2/5167/98 \\
 $ ^k$ Supported by Russian Foundation for Basic Research 
       grant no. 96-02-00019 

   } \end{flushleft}
\newpage

\section{Introduction}
The observation that the photon has a hadronic structure was first made
more than 30 years ago when the cross-sections of hadronic photoproduction
interactions were  demonstrated to have dependences on energy and
momentum transfer which were similar to those in hadron-hadron 
interactions~\cite{hadron}. With the advent of the quark-parton model, 
and subsequently
QCD, more quantitative predictions for this hadronic structure became
available, and  its gross features were identified
experimentally in $e^+e^-$ interactions~\cite{egamma}. Subsequently these
features were also observed in hard photoproduction
processes~[3-15].

\begin{figure}[htb] \centering        
\epsfig{file=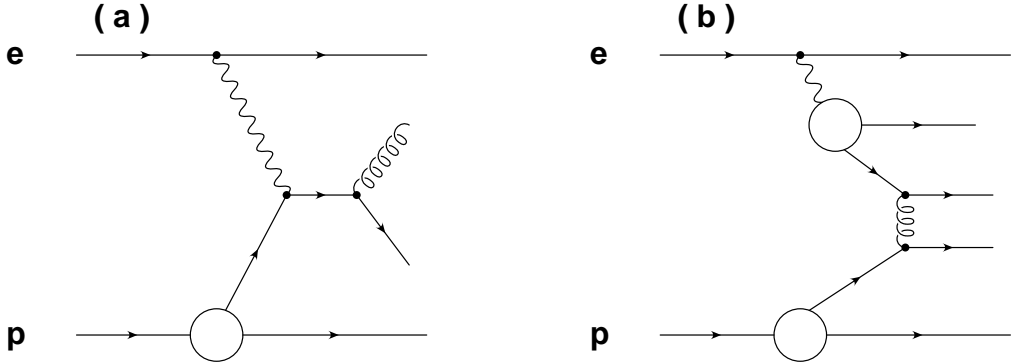,bbllx=0pt,bblly=450pt,bburx=550pt,bbury=700pt,width=150mm}
\caption{ Examples of diagrams for direct (a) and resolved (b) photon
processes in $ep$ scattering; in (b) a gluon has the space-like virtuality 
of the
probe and  a quark is resolved in the photon; similarly  gluons
in the photon  can 
be resolved by quarks and gluons with space-like virtuality.}
\label{fig1}
\end{figure}

Measurements of the photon structure functions in $e^+e^-$ interactions
are directly sensitive to the quark structure of the photon. Only through
QCD evolution studies can information be extracted concerning the gluon
component of this structure, but the presently 
available  data have not been precise
enough
for such an analysis. 
Recently, studies of jets and high $p_T$ charged particles
in  photoproduction events at the $ep$ collider HERA 
have shown sensitivity to the partonic content of the photon.
Here the photon structure is probed by the partons of the proton, rather than
by a virtual photon as in $e\gamma$ collisions. Hence these data are
sensitive
to both the quark and gluon content of the photon.  
Leading Order (LO) diagrams are shown in Fig.~\ref{fig1}, for so called 
direct (Fig.~1a; the photon couples directly to the partons in the 
proton) and resolved (Fig.~1b; the partons of the hadronic component of
the photon scatter on the partons of the proton) processes in 
photoproduction. In LO QCD 
only the latter process contains information on the 
partonic structure of the photon.

In~\cite{h1-incl} inclusive charged particle production has been studied.
It was established that the data, in particular the tail at large 
transverse momentum, $p_T$,
 can be described by Next-to-Leading Order (NLO)
QCD calculations. The charged particle 
distributions as a function of the
pseudorapidity\footnote{
$\theta$ is
the polar angle of the particle in the HERA laboratory frame,
 measured with respect to the
proton beam direction.},  $\eta   = -\mbox{ln tan}(\theta /2)$,
were found to be sensitive to the partonic structure of the 
photon. In this paper we present
 differential  cross-sections ${\rm d}\sigma/{\rm d}p_T^2$ 
for $|\eta| < 1$ in the HERA laboratory system, and
${\rm d}\sigma/{\rm d}\eta$ for $p_T
 > 2 $ GeV/c and $p_T
 > 3 $ GeV/c for charged particles, measured with the H1 detector.
 Photoproduction events have been selected by tagging the 
scattered electron.
The data  are based on an
event  sample which is
50 times larger than the one used in  \cite{h1-incl}.

In \cite{h1_gluon} a first measurement of the gluon content of the photon
was presented, while in \cite{h1-4} an effective parton density
for the photon was extracted.
In this paper the distribution of the momentum fraction
$\xg = E_{parton}/E_{\gamma}$ of the parton 
of the photon entering the hard scattering
process shown in Fig.~1b 
is measured and used to extract the
gluon density in leading order 
in the photon. 
This analysis follows closely that 
presented in \cite{h1_gluon}, where the kinematics of the 
hard scattering process were reconstructed by 
means of  jets measured in the detector. 
In the  analysis presented here the measured variable
$\xgr$, which is found to be   correlated with 
the true value of $\xg$ of the parton entering the scattering,
is based on  charged tracks with a 
high transverse momentum 
$p_T$. This method thus avoids  two large systematic errors
entering in the jet analysis. These are  the
energy scale uncertainty of the calorimeter and the uncertainty of the
jet energy measurement due to overlap with energy deposits from 
soft multiple interactions which may occur  on top 
of the hard scattering process. The drawback of this method
is a stronger sensitivity
to fragmentation uncertainties.

\section{The Apparatus}
The data were collected with the multi-purpose detector H1
 at the 
HERA collider in 1994, in which electrons\footnote{In 1994 the incident 
lepton at HERA was a positron, but we keep the generic name
``electron'' for the incident and scattered lepton throughout this paper}
 of 27.5 GeV collided with protons of 820 GeV.
The total data sample corresponds to 
a luminosity of $1.35 \pm 0.02$ pb$^{-1}$.

A detailed description of the H1-detector
has been given elsewhere~\cite{H1}.
Here only the components crucial for this particular
analysis will be briefly described.

The H1 luminosity system
consists of an electron tagger and a photon tagger,
located 33~m and 103~m from the interaction point
in the electron beam direction, respectively.
The luminosity is determined from the rate of Bremsstrahlung
process
$e p \rightarrow e p \gamma$ events by detecting the  photon.
Both detectors are TlCl/TlBr crystal
\v{C}erenkov calorimeters with an energy resolution of $10 \% / \sqrt{E
(\mbox{GeV})}$.
The electron detector allows to tag the  photoproduction
events used in this analysis
by detecting electrons scattered at small angles.

The central tracker (CT) 
consists of inner and outer cylindrical jet chambers, $z$-drift 
chambers and proportional chambers.
The two
cylindrical drift chambers~\cite{cjc} (CJC),
are mounted concentrically around the
beamline inside a homogeneous magnetic field of 1.15 Tesla and provide
up to 65 space points for a charged track, yielding
particle charge and momentum from the track curvature
in the polar angular range of
20$^{\circ}$ and 160$^{\circ}$.
The transverse momentum resolution is 
$\sigma (p_T) / p_T \sim 0.6\%\cdot p_T~(\mbox{GeV/c})$ in the  $p_T$ range
$2 < p_T < 12$ GeV/c considered here.
 The resolution for the polar angle $\theta$ is 2~mrad.

The central tracking system is complemented by a forward tracking
system. All trackers are
surrounded by a fine grained liquid argon sampling
calorimeter, consisting of an electromagnetic section
with lead absorbers and a hadronic section with steel absorbers.

\section{Data Analysis}

\subsection{Event Selection}

Photoproduction events were selected which have the scattered electron 
detected in the electron tagger, and consequently have
a four-momentum transfer $Q^2 < 10^{-2}$ GeV$^2$, where
$Q^2 = 4E_e E_{e}' \mbox{cos}^2 (\theta /2)$. 
Here $E_e$ and $E_{e}'$ are the incoming and scattered electron
energy and $\theta$ the polar angle of the scattered electron.
Tagged events were
 required to have an electron candidate in the fiducial volume
of the electron-tagger with energy $E_{e}'> 4~\mbox{GeV}$ and to have less than
2 GeV deposited in the photon detector.  
The latter condition suppresses background
from the proton beam, which appears in a random
coincidence with the high rate Bethe-Heitler events, and also reduces 
 QED corrections. 
In addition a track-based trigger was required, which is formed using only 10
selected layers out of 56 radial signal wire layers of the CJC. It demands
the presence of at least one  track of negative charge with
$p_T~>~700~\mbox{MeV/c}$ and with a distance of closest approach
of less than 2 cm from the nominal beam axis; this
requirement suppresses beam-wall background as well as fake tracks 
from random hits due to synchrotron  radiation.

For this analysis the events were required to have
 $0.3<y=E_{\gamma}/E_e< 0.7$ where $E_{\gamma}$
 is the photon  energy.  
As a consequence, in the 
$\gamma p$ centre-of-mass system (CMS) the energy
range 
is $165 <\sqrt{s_{\gamma p}}<251~\mbox{GeV}$, 
with an average of $\sqrt{s_{\gamma p}} \sim 200~\mbox{GeV}$.
An interaction vertex, reconstructed using tracks in the CT, 
had to be in the range  $-25 < z_V < +35~\mbox{cm}$, where $z_V$ is the 
$z$ coordinate of the reconstructed vertex.

A total of about $960\,000$ events satisfy all described criteria. 
The remaining non-$ep$
background was determined by analysing data from electron 
pilot bunches (i.e. electron bunches which have no colliding proton
bunch partner) 
and from a monitoring trigger of minimum bias events. 
It was found to be less than $2.7\%$  at this stage of selection, 
before tighter cuts on track quality and transverse momentum were applied.

\subsection{Track Selection}

For this analysis tracks from the central 
tracker were selected which have 
a transverse momentum with respect
to  the beam axis of at least $ 2$ GeV/c, 
a pseudorapidity measured in the 
HERA laboratory reference system of 
$|\eta | < 1$ 
and a minimum track length $\Delta R$ of 30 cm 
in the $x~-~y$ plane. The latter  cut is essential to ensure a good 
measurement of $p_T$ of the track, from 2 up to 12~GeV/c, which is the 
range considered here.
To minimize the systematic errors, sectors of the CT in
$\phi$, the azimuthal angle,
which had
a lower efficiency have been excluded from the analysis.
A total of 
16591 tracks was selected for the analysis, resulting from 15543
events.

\subsection{Corrections and Background}
The residual background from beam-gas, beam-wall, and pile-up events, 
after demanding a well-measured, high $p_T$ track such as defined above, 
was found to be negligible.
 The background from cosmic events was removed by rejecting events
with two
oppositely charged tracks meeting the requirements above (except for 
$\phi$ region cut), and which were aligned in $\theta$ and $\phi$ to better 
than $5^o$. All events rejected by this cut were visually checked 
and confirmed to be cosmic 
events, and all were found to have  at least one track with $p_T 
> 6~\mbox{GeV/c}$. Conversely, all events with at least one track with 
$p_T > 6~\mbox{GeV/c}$ were visually scanned but no further cosmic 
event was found. 
A check for the  potential contamination from 
DIS overlap events was also performed: no such events were 
found.

In order to obtain the produced number of tracks, $N_{tr}$, 
the observed number of events and  tracks were corrected as follows:

\begin{itemize}
\item 
$A_{etag}$: The geometrical acceptance of the electron tagger,
determined as described in~\cite{sigma_tot}, 
was corrected as a function of $y$. On average the acceptance is
  $55\%$  for 
$0.3<y<0.7$.  The tagger efficiency within the 
chosen acceptance is $100\%$.
\item 
$\epsilon_{trig}$: The 
CJC trigger efficiency has been estimated by comparing
the detection rates for samples of events with the main trigger
and a minimum bias trigger for
$e$-tagged events. For the high $p_T$ range considered here, 
it was estimated to be $96.3 \pm 1.2\%$, independent of $p_T$, $\eta$ and
$\phi$.
\item 
$\epsilon_{track}$~: The overall track efficiency contains 
three multiplicative factors~:
(i) The efficiency for finding and reconstructing single tracks, 
$\epsilon_{rec}$, was taken from Monte Carlo studies  and found to be 
$99.2 \pm 0.8\%$. The agreement between data and Monte Carlo was
verified by visual scanning of events in both samples.
(ii) The efficiency of the cut on transverse track length, 
$\epsilon_{\Delta R}$, was estimated from data to be $97.9 \pm 0.4\%$.
This was checked to be independent of both $\eta$ and $\phi$.
(iii) The $\phi$-restriction, $\epsilon_{\phi}$, of the geometrical
acceptance of the central drift chamber was accounted for assuming a 
uniform
distribution in $\phi$, and amounts to $53.1\%$.
\item
The bin-to-bin migration in $p_T $, $\epsilon_{mig}$,
 due to the rapidly decreasing $p_T$ spectrum and to the finite resolution 
in $p_T$, was estimated from a Monte Carlo 
simulation and for the varying 
bin sizes used here it was  found to be a simple multiplicative
factor $103.8 \pm 1.3\%$, independent of $p_T$.
\end{itemize}

The total correction factor for tracks is then $51.5~\pm 2.0\%$, where the
systematic errors have been added in quadrature. 
The uncertainties on these correction factors, 
 were taken as systematic
errors.
Detailed checks and comparisons with Monte Carlo led 
to the introduction 
of  an extra
systematic error, to account for a remaining 
uncertainty in the $p_T$ measurement,
which showed up in two ways: in the ratio of positively and
negatively charged tracks  and in the efficiency determination for
 a stronger cut on the  transverse track length  ($\Delta R>50$ cm).
The
resulting uncertainty 
 was parameterized as 
 $2\%\cdot p_T$/(GeV/c) at the centre of each bin for $p_T$
distributions, and as $2\%\cdot p_T^{low}$/(GeV/c)
 for $\eta$ distributions, with 
$p_T^{low}$ the lowest $p_T$ value of particles
 included in the distribution.

All systematic uncertainties were added in quadrature, except for 
the overall uncertainty of $5\%$ from the luminosity measurement and 
the electron tagger acceptance
which  is  not included 
in the error bars shown in the figures below.

\section{Monte Carlo Simulation Programs}

The PYTHIA 5.7 event generator~\cite{pythia}
was used
to simulate photon-proton interactions.
As well as the leading order cross-section,
PYTHIA includes initial- and final-state QCD parton radiation effects
which are calculated in the leading logarithmic approximation.
The strong coupling constant
$\alpha_s$ was calculated in first order QCD using
$\Lambda_{\rm QCD}=200$\,MeV for 4 quark flavours.
For the parton distributions of the proton the  GRV-LO~\cite{pgrv} set was 
used.

The model calculations were made for different sets of the 
parton densities in the photon: GRV~\cite{ggrv}, 
SAS1D~\cite{sas} and LAC1~\cite{lac1}. 
All these parametrizations were extracted from 
QCD fits to photon structure function data measured in $e\gamma$
collisions from $e^+e^-$ interactions, 
but have different assumptions for the gluon content,
which is only weakly constrained by these measurements. In particular
LAC1 assumes a large gluon content of the photon at small-$x$ values
which is
larger than for both other parametrizations.
The GRV and  SAS1D distributions both start the evolution from 
a small starting scale, $Q^2_0 = 0.25$ GeV$^2$ and 0.36 GeV$^2$
respectively, a procedure which has turned out to be quite
successful for the parton densities in the proton. The different 
treatment of the vector meson valence quark 
distributions leads to a larger
gluon component at small-$x$ of the photon for the GRV compared to 
the SAS1D parton densities.

Multiple parton interactions were generated in
addition to the primary parton--parton scattering.
They are calculated as leading order QCD processes between partons from the
photon and proton remnants.
The
transverse momentum
of all (primary and multiple) 
parton--parton interactions was required to be above
a cut-off value of $p_T^{min}$, depending on the photon 
structure function. For the parton densities used here the 
values $p_T^{min}=1.2, 1.0$ and 2.0 GeV/c have been used for 
GRV, SAS1D and LAC1 respectively.
These  values have been found to give an
optimal description of the transverse energy
flow outside  jets~\cite{h1-3}.

The PHOJET 1.06 event generator~\cite{phojet}, based on the two-component
Dual Parton model~\cite{dpm} has been used as well. PHOJET incorporates
very detailed simulations of both multiple soft and hard parton interactions
on the basis of a unitarization scheme. It also includes initial- and 
final-state interactions. For the distributions shown
below, it was verified that the PYTHIA and PHOJET predictions for the 
GRV parton distributions for the proton and the photon agree to better than
5\%.

Hadronization in both PHOJET and PYTHIA was modelled with the LUND string
fragmentation scheme (JETSET 7.4~\cite{jetset}).
Since the kinematics of the hard scattering process will be 
related to the $p_T$ values of the particles with high transverse
momentum, it is imperative to study the fragmentation dependence of 
this measurement. A model with a different hadronization scheme, 
HERWIG~\cite{HERWIG}, was used for this purpose. In this model the 
hadronization is based on cluster fragmentation. Other parameters, such as
the parton densities for the photon and proton, were
 taken to be  the same as for
PYTHIA. Both the LUND model and HERWIG were found to give a good description
of the general features of fragmentation as measured in $e^+e^-$ 
collisions at LEP (e.g. \cite{barate}).

\section{Inclusive Charged Particle Cross Sections}

The invariant cross-section for single particle production is
given by
\begin{equation}
\frac{ {\rm d}^2\sigma} { {\rm d}p_T^2\ {\rm d}\eta }\
= \int \frac{ {\rm d}^3\sigma} { {\rm d}p_T^2\
   {\rm d}\eta\, {\rm d}\phi } \ \cdot {\rm d}\phi
= \pi \cdot E \cdot \frac{ {\rm d}^3 \sigma}{ {\rm d}p^3}\,
\end{equation}
assuming azimuthal symmetry of the cross-section
 allowing  integration over $\phi$.
The measurement was made at a $\sqrt{s_{\gamma p}} \sim 200$ GeV, 
and effectively averages over the region
 $165~\mbox{GeV}<\sqrt{s_{\gamma p}}<251~\mbox{GeV}$. No significant 
$\sqrt{s_{\gamma p}}$
 dependence was found when the data were subdivided in two 
$\sqrt{s_{\gamma p}}$ bins.
The cross-section for inclusive charged particle production
in $\gamma p$ collisions was calculated from the
corrected number of tracks produced,
$N_{tr}$, in a bin of $p_T$ and $\eta$.
It is given by:
\begin{equation}
\frac{ {\rm d}^2\sigma_{\gamma p}} { {\rm d}p_T^2\ {\rm d}\eta } =
\frac{N_{tr}(p_T,\eta) }
  { {\cal L} \cdot F \cdot 2  p_T  \Delta p_T \cdot \Delta \eta   }
\end{equation}
where  ${\cal L}$ denotes the integrated luminosity, $F$ is the
 photon flux integral and the flux $f(y)$
is defined according to
$\ \ {\rm d}\sigma(ep) =\
 \sigma(\gamma p) \cdot f(y) \cdot {\rm d}y$.
For the chosen $y$-range
the integral over $y$  of the photon flux yields $F = 0.0136$,
assuming the Weizs\"acker-Williams approximation~\cite{wwa,frixione}.
$\Delta \eta$ and
$\Delta p_T^2 \ = \ 2 \cdot p_T \cdot \Delta p_T $
are the bin widths.

\begin{figure}[thb] \centering        
\epsfig{file=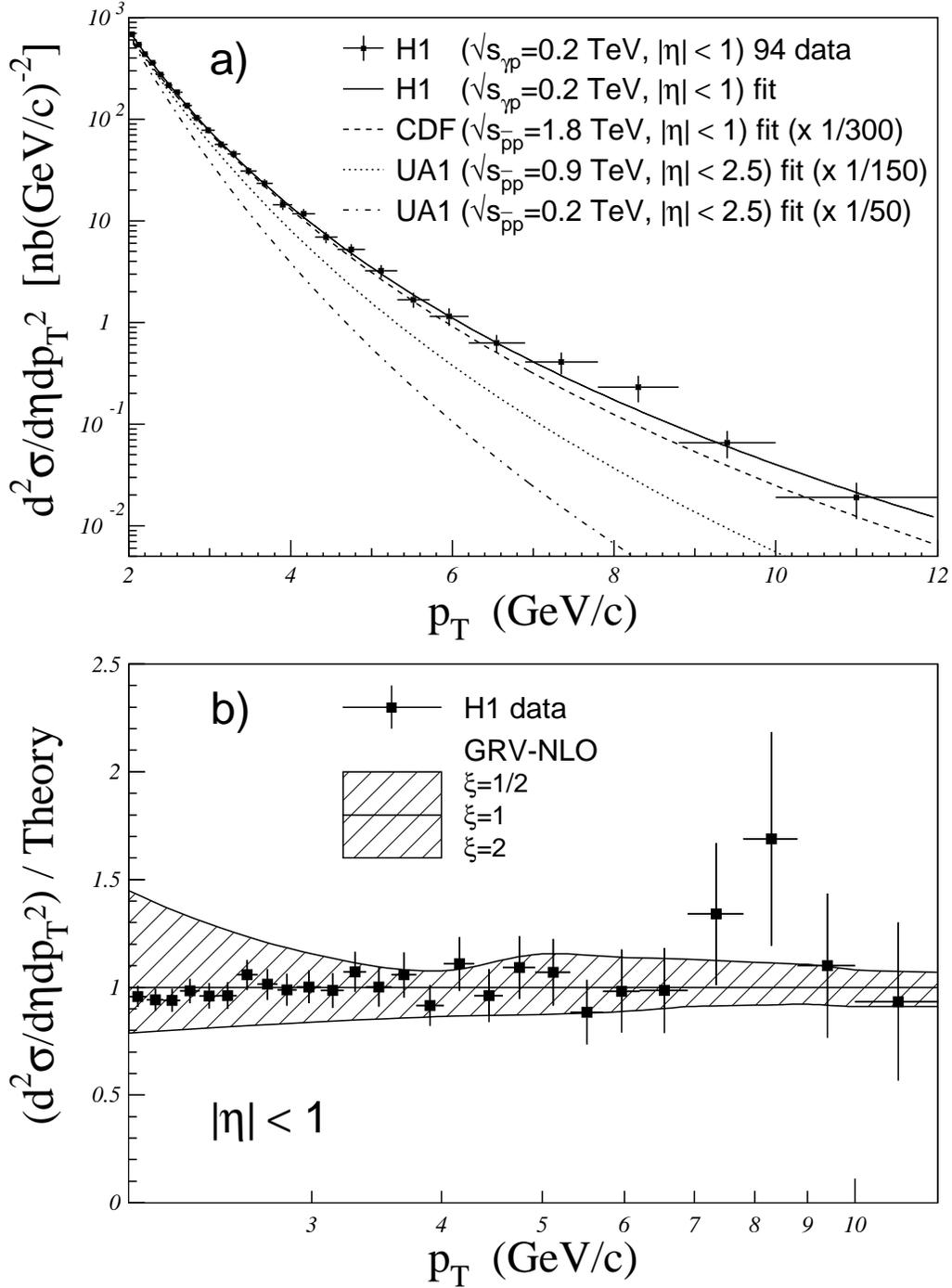,bbllx=0pt,bblly=70pt,bburx=600pt,bbury=770pt,height=19cm}
\caption{ {\bf a)} The inclusive $\gamma p$
  cross-section for charged particles in photoproduction (full squares)
  measured in the kinematical region
  $\mid \eta \mid < 1.0 $
  at an average $ \sqrt{s_{\gamma p}}\approx 200$ GeV.
  The error bars denote  the  statistical and
  systematic errors added in quadrature.
  An overall uncertainty of
  5\%  from the luminosity measurement and the electron tagger
acceptance is not included in the errors.
The curves indicate  power-law
  fits, as described in the text, for these data and for the 
$p\overline{p}$ data from UA1 and CDF as given in 
\protect \cite{ua1-1,cdf}.
{\bf b)}
 The ratio of data over the NLO QCD calculation with scale $\xi p_T^2$
for $\xi=1$.  The shaded band shows the expected variation of the ratio as 
$\xi$ changes from 0.5 to 2, illustrating the sensitivity of the QCD 
calculation to the renormalization and factorization 
scales (see text).}
\label{pt1}
\end{figure}

\begin{figure}[htb] \centering
\epsfig{file=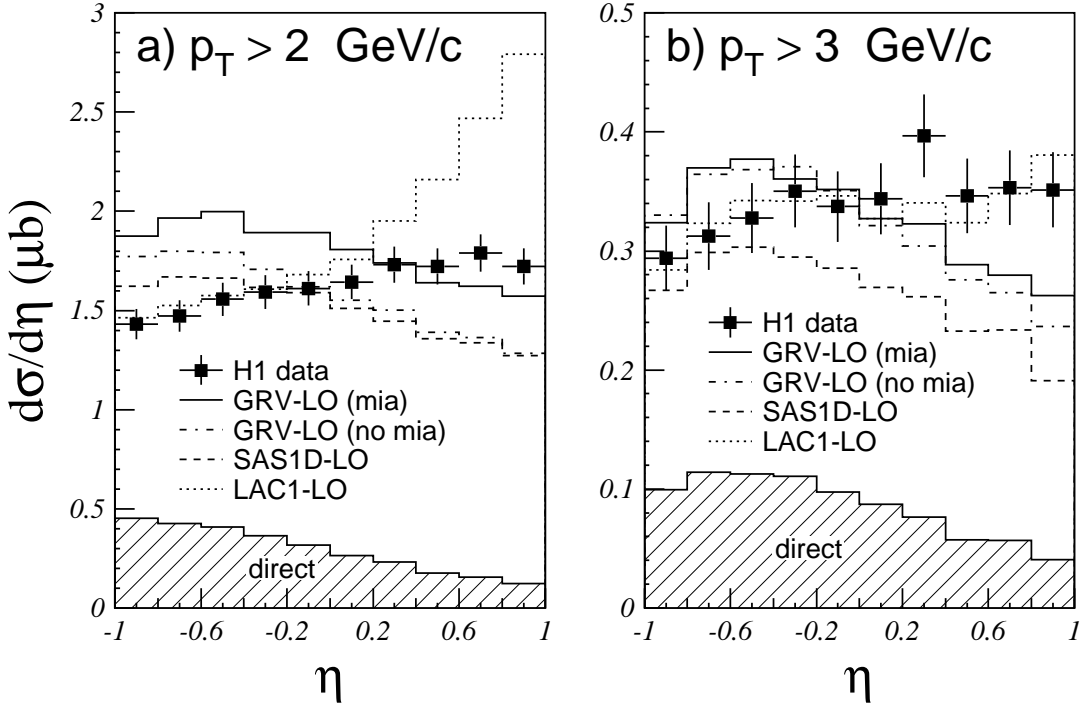,bbllx=50pt,bblly=250pt,bburx=600pt,bbury=590pt,width=17cm}
\caption{The differential $\gamma p$ cross-section 
${\rm d}\sigma/{\rm d}\eta$
         for inclusive production of  high $p_T$ charged particles
(full squares)
         in comparison with LO QCD calculation by PYTHIA (histograms).
  The error bars denote  the  statistical and
  systematic errors added in quadrature.
An overall
uncertainty of 5\% from the luminosity 
measurement and electron tagger acceptance 
is not included.
         Different lines represent different photon structure function 
         parametrizations: GRV with (full) and without (dash-dotted)
         multiple interactions, SAS1D (dashed) and LAC1 (dotted).
         The contributions from the direct photon processes are shown
         as shaded histograms.}
\label{etaMC}
\end{figure}

\begin{figure}[htb] \centering        
\epsfig{file=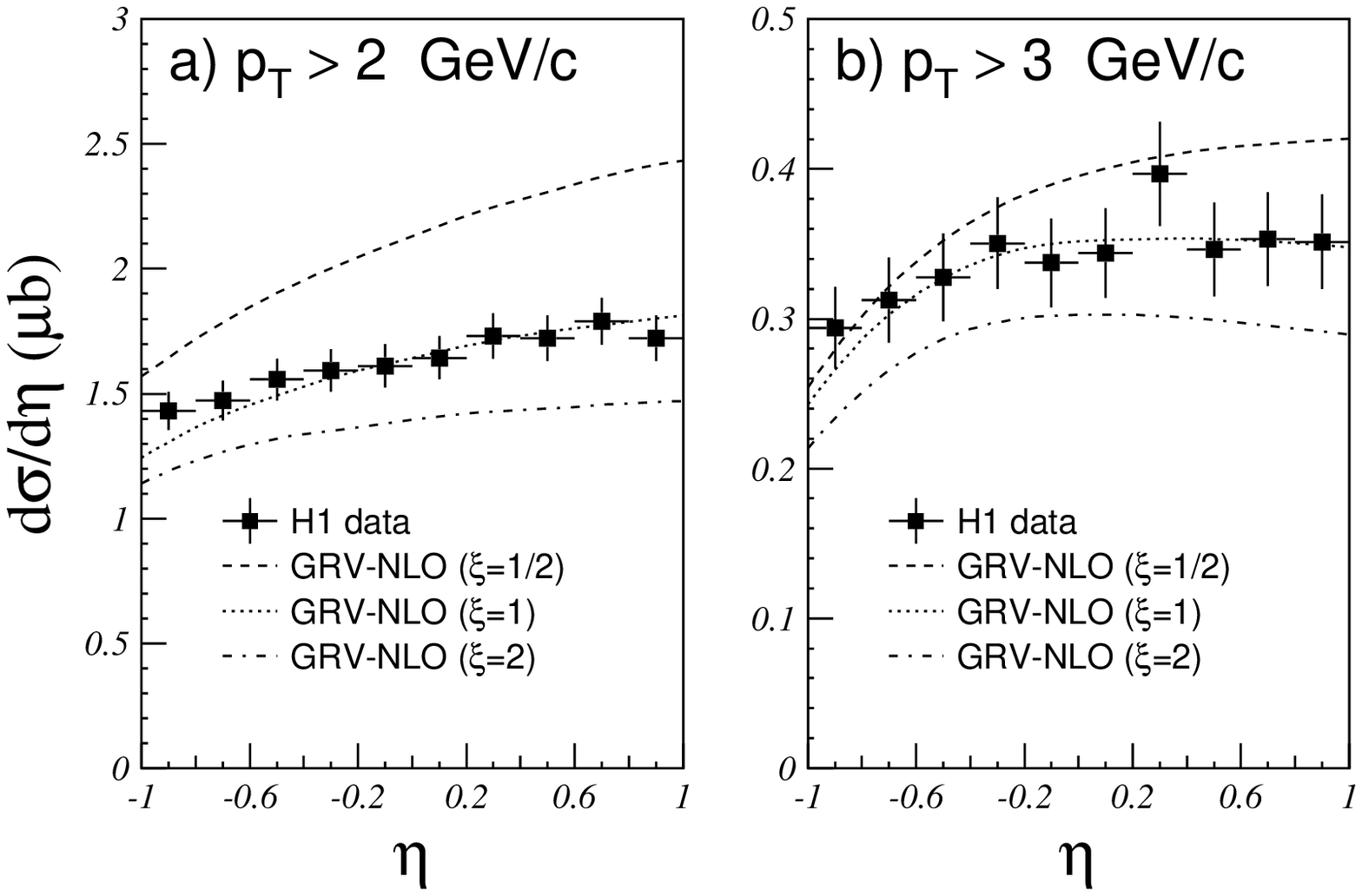,bbllx=50pt,bblly=250pt,bburx=600pt,bbury=580pt,width=17cm}
\caption{The differential $\gamma p$ cross-section ${\rm d}\sigma/{\rm d}\eta$
         for inclusive production of the high $p_T$ charged particles
(full squares)
         in comparison with NLO QCD calculations
\protect \cite{kniehl}, with different values of the 
scale, taken as $\mu = \xi p_T^2$.
  The error bars denote  the  statistical and
  systematic errors added in quadrature.
An overall 
uncertainty of 5\% from the luminosity 
measurement and electron tagger acceptance 
is not included.
 }
\label{eta1}
\end{figure}

The resulting differential cross-section
for the sum of positive and negative charged particles
is shown in Fig.~\ref{pt1}a.
  The error bars denote  the  statistical and
  systematic errors added in quadrature.
The measured cross-sections are
listed in Table \ref{tabpt}.
The data exhibit a strong improvement in precision compared
to the previous H1 measurement~\cite{h1-incl}
and cross-sections are now given up to 12 GeV/c in $p_T$.
The QCD inspired power-law~\cite{hagedorn} 
expression
\begin{equation}
 E \cdot\frac{ {\rm d}^3 \sigma} { {\rm d}p^3} =
 A \cdot (1+\frac{p_T}{(p_T)_0})^{-n}\label{fit}.
\end{equation}
was fitted to the data.
The fitted curve is shown in Fig.~\ref{pt1}a, and describes the data well
over the whole $p_T$ range.
The fit gives $A=5.44 \pm 0.66$ mb and $n=7.03 \pm  0.07$ (statistical
errors). 
The parameter $(p_T)_0$ was fixed to the value 0.63, as found in 
\cite{h1-incl}, but other values in the range 0.5--0.75 were found 
to give equally good fits. The effect of the choice of $(p_T)_0$
leads to an uncertainty in the power $n$ of about 0.2.

Also shown in Fig.~\ref{pt1}a are the
results  of similar fits to 
the $p{\bar p}$ collider data of the UA1-collaboration~\cite{ua1-1} 
in the
rapidity
region $ \mid \eta \mid < 2.5$, at CMS energies
 $\sqrt{s}=200$ and 900 GeV, and  of CDF~\cite{cdf}
in the rapidity region $ \mid \eta \mid < 1 $, at $ \sqrt{s}=1800$  GeV, 
scaled by the factors indicated in the figure, which essentially
normalizes all data to the $\gamma p$ cross-section at $p_T = 2$~GeV/c.
The high $p_T$ tail is observed to increase with increasing 
energy in $p{\bar p}$. The hardness of the $\gamma p $ spectrum
is comparable to that of  
the  $p{\bar p}$ data  at the highest energy. 
The high $p_T$ tail in the data is clearly 
larger than in $p\overline{p}$
collisions at similar  CMS energy, which
can be understood as
being due to extra contributions in $\gamma p$, namely the 
direct and the pointlike resolved (or anomalous) component~\cite{h1-incl}.

\begin{table}
\begin{center}
\begin{tabular}[hbpt]{|c|c|c|}
\hline 
& & \\[-0.35cm]
$p_T$ [GeV/c]&   $\frac{{\rm d}\sigma^{\gamma p}}{{\rm d}p_T^2{\rm d}\eta}$ 
(nb/(GeV/c)$^2$)  &
$\Delta \frac{{\rm d}\sigma^{\gamma p}}{{\rm d}p_T^2{\rm d}\eta}$ (nb/(GeV/c)$^2$) \\
& & \\[-0.35cm]
\hline 
 2.00 -- 2.08  &   685.0    &     35.2 \\
 2.08 -- 2.16  &   539.1    &     29.3 \\
 2.16 -- 2.24  &   434.2    &     24.7 \\
 2.24 -- 2.34  &   359.9     &      20.7   \\
 2.34 -- 2.44  &   274.3     &      16.7   \\
 2.44 -- 2.54  &   216.6     &      13.9   \\
 2.54 -- 2.66  &   184.8     &      12.1   \\
 2.66 -- 2.78  &   135.9     &       9.53  \\
 2.78 -- 2.90  &   102.8     &       7.64  \\
 2.90 -- 3.06  &    78.32    &       5.97  \\
 3.06 -- 3.22  &    56.41    &       4.59  \\
 3.22 -- 3.38  &    45.62    &       3.97  \\
 3.38 -- 3.58  &    31.01    &       2.82  \\
 3.58 -- 3.78  &    23.47    &       2.32  \\
 3.78 -- 4.02  &    14.33    &       1.49  \\ 
 4.02 -- 4.30  &    11.79    &       1.33  \\
 4.30 -- 4.58  &     6.905   &       0.885 \\
 4.58 -- 4.92  &     5.210   &       0.692 \\
 4.92 -- 5.32  &     3.224   &       0.465 \\
 5.32 -- 5.72  &     1.673   &       0.283 \\
 5.72 -- 6.20  &     1.148   &       0.225 \\
 6.20 -- 6.90  &     0.631   &       0.126 \\
 6.90 -- 7.80  &     0.408   &       0.100 \\
 7.80 -- 8.80  &    0.232    &       0.068 \\
 8.80 -- 10.0  &    0.066    &       0.020 \\
 10.0 -- 12.0  &    0.019    &       0.007 \\
\hline      
\end{tabular}
\caption{ Measured differential cross-section and total errors
$(\Delta)$ for the production 
of charged particles in the $\eta$ range $-1$ to $1$. An overall 
uncertainty of 5\% from the luminosity and electron tagger acceptance 
is not included.}

  \label{tabpt}
\end{center}
\end{table}

\begin{table}
\begin{center}
\begin{tabular}[hbpt]{|c|c|c|c|c|}
\hline
 & \multicolumn{2}{c|}{$p_T> 2.0$ GeV/c}
 & \multicolumn{2}{c|}{$p_T> 3.0$ GeV/c}\\
\hline
& & & & \\[-0.35cm]
  $\eta$  &  $\frac{{\rm d}\sigma}{{\rm d}\eta}$ (nb)  &  
$\Delta\frac{{\rm d}\sigma}{{\rm d}\eta}$ (nb) & 
$\frac{{\rm d}\sigma}{{\rm d}\eta}$ (nb)  &  
$\Delta\frac{{\rm d}\sigma}{{\rm d}\eta}$ (nb) \\
& & & & \\[-0.35cm]
\hline 
  $-1.0$ -- $-0.8$  &       1431.1     &      77.8 & 293.9     &       27.2 \\
  $-0.8$ -- $-0.6$  &      1473.1      &      79.4 & 312.3     &       28.4 \\
  $-0.6$ -- $-0.4$  &      1556.3      &      83.2 & 327.6     &       29.3 \\
  $-0.4$ -- $-0.2$  &      1593.3      &      85.1 & 350.3     &       30.7 \\
  $-0.2$ -- $-0.0$  &      1611.9      &      85.3 & 337.4     &       29.6 \\
  0.0 -- 0.2  &      1643.8      &      87.5   & 343.7     &       29.8 \\
  0.2 -- 0.4  &      1730.2      &      91.2   & 396.8     &       34.8 \\ 
  0.4 -- 0.6  &      1721.3      &      90.9   & 346.1     &       31.4 \\
  0.6 -- 0.8  &      1790.4      &      94.0   & 353.3     &       31.3 \\
  0.8 -- 1.0  &      1721.2      &      90.6   & 351.4     &       31.5 \\
\hline      
\end{tabular}
\caption{ Measured differential cross-section and total errors
$(\Delta)$ for the production 
of charged particles. An overall 
uncertainty of 5\% from the luminosity and electron tagger acceptance 
is not included. }
  \label{tabeta2}
\end{center}
\end{table}

In Fig.~\ref{pt1}b the ratio of  data to the 
NLO  calculation~\cite{kniehl}  
 using GRV structure functions for 
photon and proton is shown. 
Both renormalization and factorization scales were chosen to be 
$p_T^2$. The 
effect of this choice is shown by predictions for a scale 
of $\xi p_T^2$, with $\xi = 0.5$ and 2, shown as a ratio to the 
prediction for $\xi = 1$.
The calculations describe the data well, in particular when 
$\xi$ is close to one.

The dependence of the cross-section on the pseudorapidity $\eta$,
$ {\rm d}\sigma / {\rm d}\eta  $, is shown in Fig.~\ref{etaMC}
for all particles with $p_T$ larger than 2 and $3$ GeV/c.
The measured cross-sections are listed in Table~\ref{tabeta2}.
Note, that the $\eta$ measured in the HERA laboratory frame
is on average shifted by +2 units with respect to the
$\eta^{*}$ distribution in
the $\gamma p$ CMS system. The spread of
this shift due to the photon energy range is less than 0.3 units
in pseudorapidity. Therefore the $\eta$ and $\eta^{*}$
distributions look very similar.
The data are consistent with being flat in most of the region, with 
a slight decrease towards $\eta = -1$, i.e. in the photon
direction. 
The data are compared with predictions of  LO QCD calculations made 
with PYTHIA, for different structure functions of the photon and 
assumptions on the soft interactions in the underlying event
(cf. section 4) .
Predictions are shown for  
 GRV with (full) and without (dash-dotted)
         multiple interactions, SAS1D (dashed) and LAC1 (dotted).
For the latter two predictions multiple interactions were 
included.
All structure functions except  LAC1 show a falling distribution with 
increasing $\eta$, contrary to the data.
        The contributions from the direct photon processes
are shown
         as shaded histogram and decrease towards
the proton direction.
For the data with $p_T> 2$ GeV/c, there is a strong sensitivity to the 
parton distributions in the photon, and also to the amount of 
underlying interactions as shown by the GRV curves.
None of the predictions presented shows  satisfactory agreement with the
data. For data with $p_T > 3 $ GeV/c, the effect of multiple 
interactions becomes very small, as shown for the GRV 
predictions. The sensitivity to the photon structure is 
also reduced but is
still significant. 
For large $p_T$ the data agree best with LAC1, and disagree with SAS1D.
It was checked that these results do not depend on the 
choice of the parton distributions of the  proton, when selected 
from those consistent with the most recent proton structure function data.

In Fig.~\ref{eta1} the cross-sections as 
a function of $\eta$ are compared with
NLO calculations, including direct and resolved contributions.
The NLO calculations~\cite{kniehl} reproduce the data, however
one can see a significant effect of the choice of the QCD scale on both the 
shape and normalization of the distributions, as shown by the curves for 
different $\xi$ values.
The effect is  smaller for data with $p_T > 3 $ GeV/c.
The scale has also been  found to be the reason for the considerable 
difference between the LO and NLO GRV cross-section predictions.
PYTHIA uses as a scale the $p_T$ of the hard partonic interaction,
while the program of \cite{kniehl} uses the $p_T$ of the final state
particle, which can  differ substantially  from
that of the original
parton.

\section{The {\boldmath $\xg$ } Distribution and Gluon Density in the Photon}

In this section the distribution of the momentum fraction
$\xg = E_{parton}/E_{\gamma}$ of the parton 
of the photon entering the hard scattering process will be determined
from the charged tracks.
For this analysis events were kept which have at least one track
reconstructed with a $p_T > 2.6$ GeV/c. This value is a compromise between
large systematic 
uncertainties  from multiple interactions and the statistical 
precision.
In total 9378 events have been selected.

For each event with at least one measured track of a 
charged particle with $p_T> 2.6 $ GeV/c the  variable 
$\xgr$ was calculated: 
\begin{equation}
\xgr = \frac{\sum p_T e^{-\eta}}{E_{\gamma}}
\end{equation}
where the sum runs over all tracks with  $p_T > 2$ GeV/c.
This variable is correlated to the true $\xg$ of the interaction.
Apart from the strongly reduced sensitivity to 
 effects of multiple interactions,
the requirement $p_T > 2.6$ GeV/c ensures that 
the data stay safely away from the 
region which is affected by the 
$p_T^{min}$ cut used in the Monte Carlo event generation
(cf. section 4).

\begin{figure}[htb] \centering        
\epsfig{file=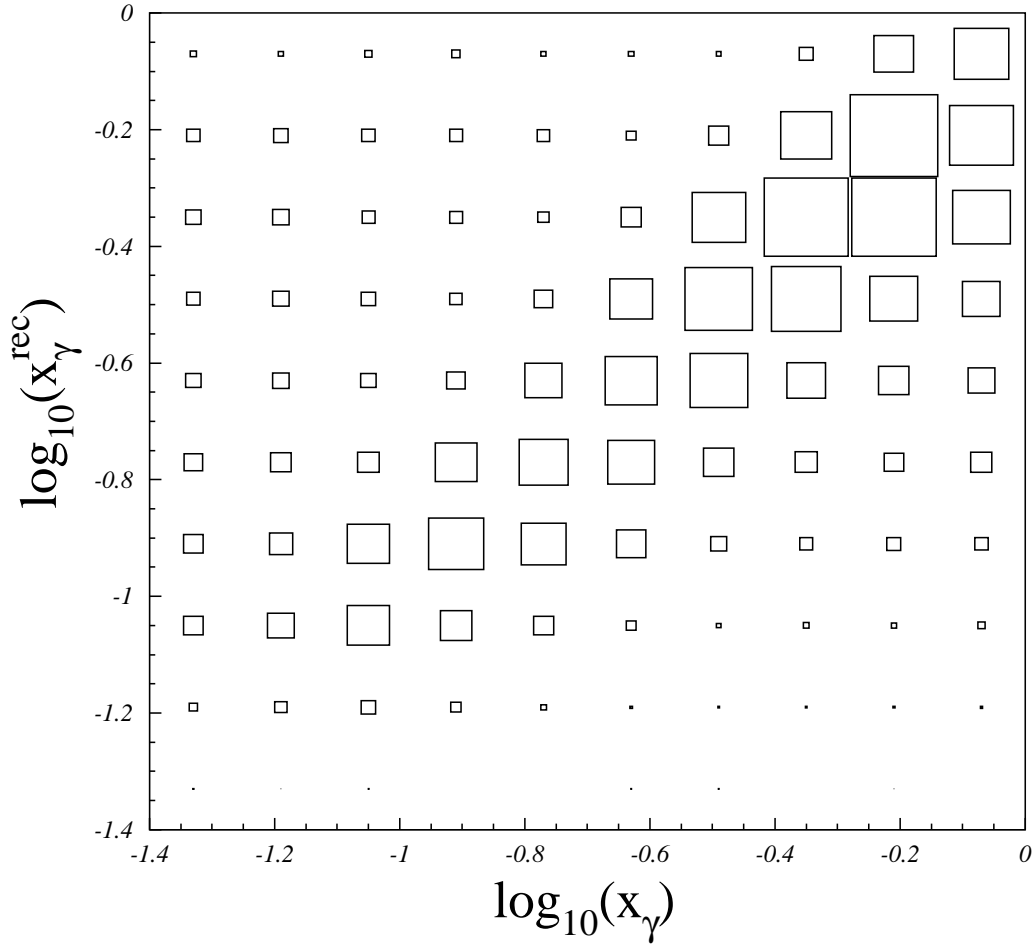,bbllx=30pt,bblly=0pt,bburx=580pt,bbury=500pt,width=150mm}
\caption{ Correlation between reconstructed and true $\xg$ values
using the PYTHIA Monte Carlo with the
LAC1 parton densities for the photon.}
\label{fig2}
\end{figure}

Monte Carlo events were  used to study the correlation between 
the measured $\xgr$ and 
the true $\xg$, shown in Fig.~\ref{fig2}. The 
generated events were fed into the detailed H1 simulation program 
and then subjected to the same reconstruction and analysis chain as 
the real data. The correlation is shown for the Monte Carlo events 
generated with the PYTHIA program, using the LAC1 
parametrization of  the structure function
of the photon.

\begin{figure}[htb] \centering        
\begin{picture}(0,0)
\end{picture}
\begin{picture}(0,0)  \end{picture}
\epsfig{file=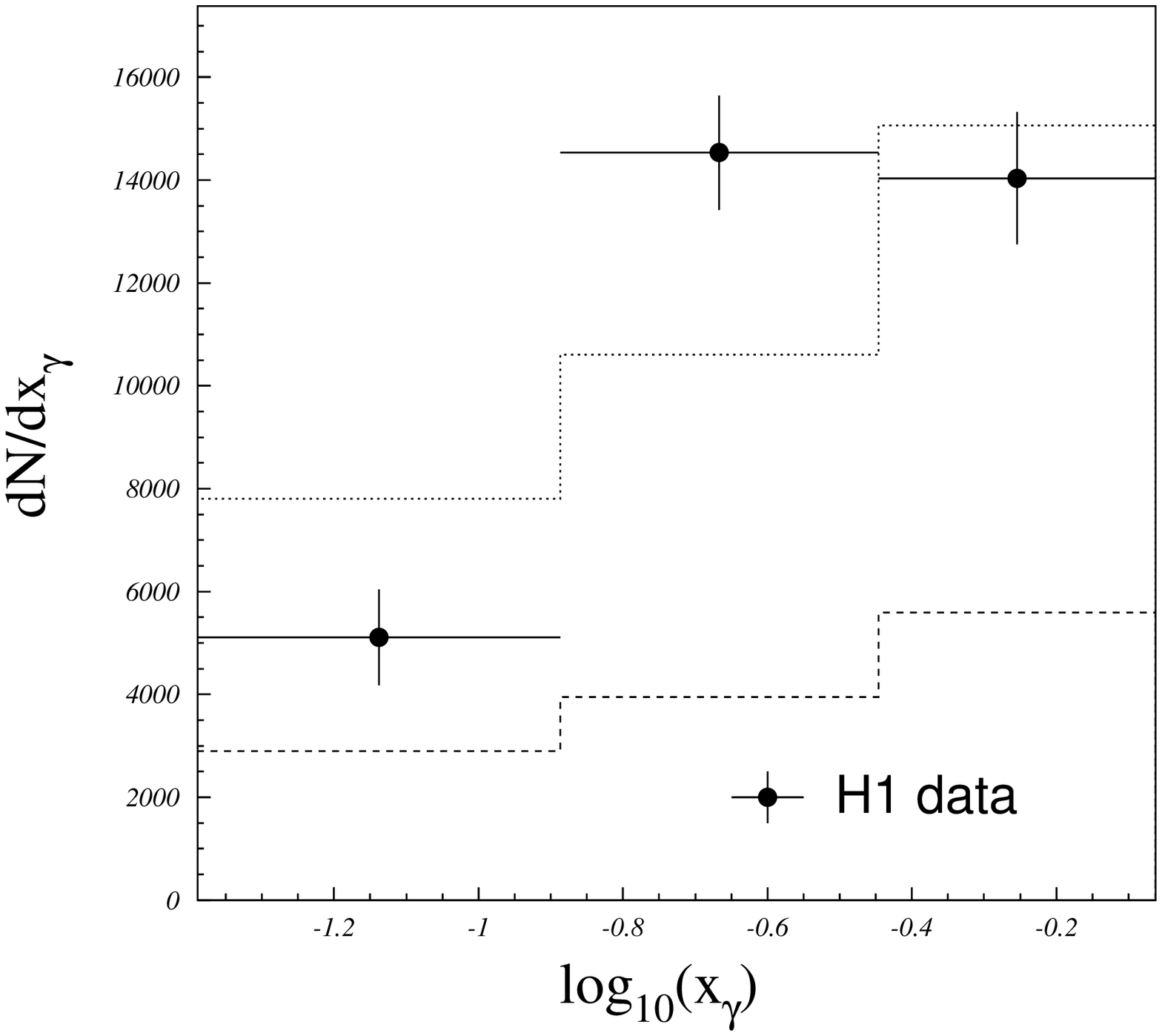,bbllx=30pt,bblly=0pt,bburx=580pt,bbury=500pt,width=150mm}
\caption{ The  
$\xg$ distribution in the photon. The errors are statistical only.
The curves are full (dotted line) and `quark only' 
(dashed line) predictions of PYTHIA, using the 
LAC1 parton distributions of the photon.
}\label{fig3}
\end{figure}

For the reconstruction of the true $\xg$ 
an unfolding procedure was used~\cite{blobel},
following the analysis as detailed in~\cite{h1_gluon}. 
Monte Carlo events generated with PYTHIA  using the LAC1 
parton densities have been used for the unfolding.
The result is presented in three bins in $\xg$, a condition
 imposed by the 
unfolding procedure, in order to minimize the bin-to-bin correlations.
The resulting
$\xg$ distribution  is shown 
in Fig.~\ref{fig3}, with statistical errors only, which include the 
bin-to-bin correlations from the unfolding.
The dotted curve is the prediction of the PYTHIA model with the LAC1 
parton distributions for the photon. The dashed curve shows the 
component of the event rate 
where only  quarks from the photon side are involved.
The quark distribution in the photon is rather well constrained
from  $e\gamma$
measurements as discussed in section 1 above.
The full calculation gives a fair description of the measurement, and a 
large contribution of gluon induced processes 
(from the photon side)~\cite{h1_gluon}
is clearly confirmed.

\begin{figure}[htb] \centering        
\begin{picture}(0,0)  \end{picture}
\epsfig{file=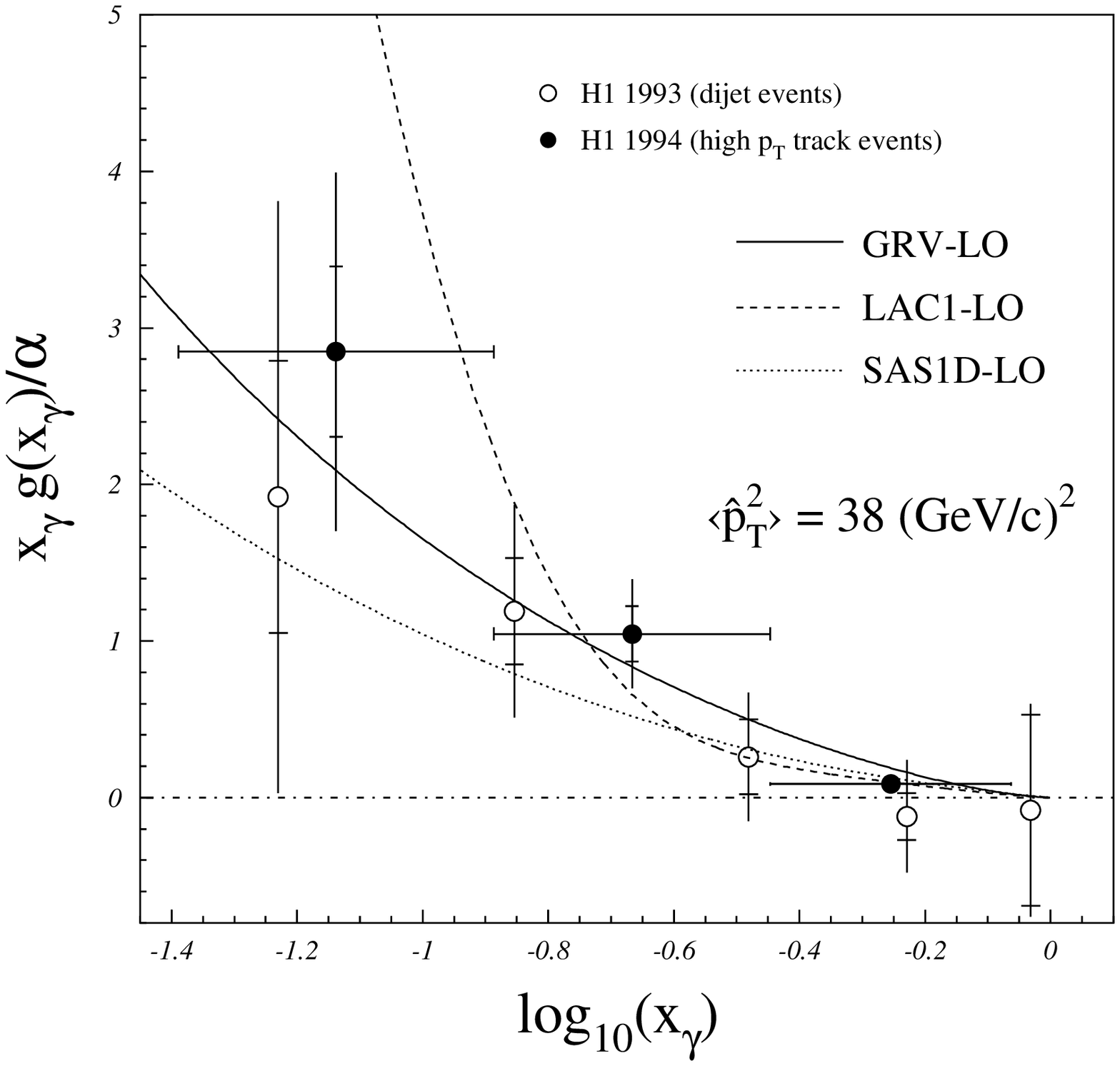,bbllx=50pt,bblly=-10pt,bburx=550pt,bbury=550pt,width=150mm}
\caption{ The  LO
gluon distribution in the photon from charged tracks
(full circles; $<p_T^2> = 38$ GeV$^2/c^2$) and 
jets (open circles; $<p_T^2> = 75$ GeV$^2/c^2$). 
The inner error bars are statistical,
the full bars are the statistical and systematic errors added in 
quadrature.   An overall uncertainty of
  5\%  from the luminosity measurement and the electron tagger
acceptance is not included in the errors.
The curves are GRV~\cite{ggrv} (full line), SAS1D~\cite{sas}
 (dotted line) and LAC1~\cite{lac1} (dashed line).}
\label{fig4}
\end{figure}

The 
gluon distribution in the photon, extracted at leading order,
was then obtained from the $\xg$
distribution  as follows. 
The contribution of processes in which quarks are resolved in the photon
(for example Fig.~\ref{fig1}(b)) were calculated and subtracted from the data.
The ratio of the subtracted data distribution with the distribution 
calculated from Monte Carlo using only
processes initiated by a gluon from the photon side  
gives 
weight factors. Applying these weight factors  to the 
input gluon distribution used in the Monte Carlo yields the 
measured  gluon distribution, which is  shown in Fig.~\ref{fig4}. 
The average transverse momentum squared  of the hard partonic scattering for 
this data sample amounts to $<p_T^2> = 38$ GeV$^2/c^2$, 
as derived from Monte Carlo studies, and is taken as 
the scale for comparisons with parton densities.
The total errors include
statistical and systematic errors where all contributions have been added
in quadrature. Apart from the systematic errors already included in the 
cross-section measurements discussed previously, 
the following systematic errors specific 
to the unfolding had to be taken into account:
\begin{itemize}
\item 
The uncertainty in the quark density of the photon was found to be 15\% 
by repeating the analysis with different parametrizations of the quark 
distributions~\cite{ggrv,sas,lac1}.
\item A variation of the unfolding parameters led to a 10\% uncertainty.
\item 
 Changes of up to 20\% were observed when different fragmentation 
models (PYTHIA, PHOJET, HERWIG) and different parton densities for the 
photon structure (GRV, LAC1)
were selected in the Monte Carlo used for the unfolding.  
Hence a 20\% systematic error from this source was assumed.
\end{itemize}
The uncertainty in the parton densities of the proton is negligible 
compared to the uncertainties above and this was neglected.

The result is compared with the measurements based on the 
1993 H1 data~\cite{h1_gluon}, using jets
instead of charged tracks. 
The average transverse momentum squared  of the hard partonic scattering for 
the jet data sample amounts to $<p_T^2> = 75$ GeV$^2/c^2$.
The measurements are found 
to be consistent. The  
improvement in  precision of the measurement 
presented here is clearly visible.
The results confirm that the contribution of the gluon to  photon 
structure is significant. 
The gluon density tends to rise with decreasing $\xg$.
The result is compared with various parton distributions: LAC1,
GRV, SAS1D. They are
generally in agreement with the data.

\section{Conclusions}

The differential cross-sections ${\rm d}\sigma/{\rm d}p_T^2, $
for $|\eta| < 1$ in the HERA laboratory frame, and 
${\rm d}\sigma/{\rm d}\eta$ for 
$p_T> 2$ GeV/c and $p_T> 3$ GeV/c
have been measured in photoproduction events with the H1 detector.
The $p_T$ spectrum exhibits a high $p_T$ tail, larger than in $p\overline{p}$
collisions at similar centre of mass  energy.
The $\eta$ spectra show sensitivity to the parton densities in the photon.
When charged particles down to $p_T = 2$ GeV/c are included, there is 
also  
a significant effect from the activity in the underlying event.  
Charged tracks  have been used to extract information on the 
hadronic structure of the photon, by measuring the $\xg$ distribution.
The $\xg$ distribution has been
unfolded from the data using for the first time high $p_T$ charged
tracks to 
extract the LO gluon density of the photon.  The gluon density 
is found  to increase with decreasing $\xg$ in agreement with 
an earlier 
H1 analysis using jets to tag and reconstruct the hard sub-process.

\section*{Acknowledgements}

We are grateful to the HERA machine group whose outstanding
efforts have made and continue to make this experiment possible. 
We thank
the engineers and technicians for their work in constructing and now
maintaining the H1 detector, our funding agencies for 
financial support, the
DESY technical staff for continual assistance, 
and the DESY directorate for the
hospitality which they extend to the non-DESY 
members of the collaboration.
We have benefited from a number of helpful discussions with B.A. Kniehl.





\begin{thebibliography}{99}
\bibitem{hadron} J. J. Sakurai, Ann. Phys. 11 (1960) 1;\\
M. Gell-Mann and F. Zachariasen, Phys. Rev. 124 (1961) 953.
\bibitem{egamma}
PLUTO Collab., Ch. Berger et al.,      Nucl. Phys.     B281    (1987)  365.\\
JADE Collab., W. Bartel et al.,        Z. Phys.        C24     (1984)  231.\\
TASSO Collab., M. Althoff et al.,      Z. Phys.        C31     (1986)  527.\\
TPC/2$\gamma$ Collab., H. Aihara et al., Z. Phys.      C34     (1987)    1.\\
AMY Collab., S.K. Sahu et al.,         Phys. Lett.     B346    (1995)  208.\\
TOPAZ Collab., K. Muramatsu et al.,     Phys. Lett.     B332    (1994)  477.\\
DELPHI Collab., P. Abreu et al.,       Z. Phys.        C69     (1996)  223.\\
OPAL Collab., K. Ackerstaff et al.,    Z.    Phys.     C74     (1997)   33.\\
\bibitem{h1-1} H1 Collab., I. Abt et al., Phys. Lett. B297 (1992) 205.
\bibitem{h1-2} H1 Collab., I. Abt et al., Phys. Lett. B314 (1993) 436.
\bibitem{h1-incl} H1 Collab., I. Abt et al., Phys. Lett. B328 (1994) 176.
\bibitem{h1_gluon}
H1 Collab., T. Ahmed et al.,           Nucl. Phys.     B445    (1995)  195.
\bibitem{h1-3} H1 Collab., S. Aid et al., Z. Phys. C70 (1996) 17.
\bibitem{h1-4} H1 Collab., C. Adloff et al., Eur. Phys. J. C1 (1998) 97.
\bibitem{zeus-1} ZEUS Collab., M. Derrick et al., Phys. Lett. B297 (1992) 404.
\bibitem{zeus-2} ZEUS Collab., M. Derrick et al., Phys. Lett. B322 (1994) 287.
\bibitem{zeus-3} ZEUS Collab., M. Derrick et al., Phys. Lett. B342 (1995) 417.
\bibitem{zeus-4} ZEUS Collab., M. Derrick et al., Phys. Lett. B348 (1995) 665.
\bibitem{zeus-5} ZEUS Collab., M. Derrick et al., Z.  Phys. C67 (1995) 227.
\bibitem{zeus-6} ZEUS Collab., M. Derrick et al., Phys. Lett. B384 (1996) 401.
\bibitem{zeus-7} ZEUS Collab., J. Breitweg et al., Eur. Phys. J. C1 (1998) 109.
\bibitem{H1}
H1 Collab., I. Abt et al.,
Nucl. Instr. and Methods A386 (1997) 310. \\
H1 Collab., I. Abt et al.,
Nucl. Instr. and Methods A386 (1997) 348.
\bibitem{cjc}
J. B\"urger et al.,                  Nucl. Instr. and Methods
                                                       A279    (1989)  217.
\bibitem{sigma_tot} H1 Collab., S. Aid et al., Z. Phys. C69 (1995) 27. 
\bibitem{pythia}
T. Sj\"ostrand, CERN-TH-6488 (1992),
                                     Comp. Phys. Commun.
                                                       82      (1994)   74.
\bibitem{pgrv}
M. Gl\"uck, E. Reya and A. Vogt,     Z. Phys.          C53     (1992)  127.
\bibitem{ggrv}
M. Gl\"uck, E. Reya and A. Vogt,     Phys. Rev.        D46     (1992) 1973.
\bibitem{sas}  G. A. Schuler, T. Sjostrand, Phys. Lett. B376 (1996) 193.
\bibitem{lac1} H. Abramowicz, K. Charchula, A. Levy, Phys. Lett.
B269 (1991) 458.
\bibitem{phojet} R. Engel, Z. Phys. C66 (1995) 203.\\
R. Engel and J. Ranft, Phys. Rev. D54 (1996) 4244.
\bibitem{dpm} A. Capella et al., Phys. Rep. 236 (1994) 225.
\bibitem{jetset}
T. Sj\"ostrand, M. Bengtsson,           Comp. Phys. Commun. 43 (1987) 367.
\bibitem{HERWIG} G. Marchesini  et al., Comp. Phys. Comm. 67 (1992) 465.
\bibitem{barate} R. Barate et al., ALEPH Collab., Physics Reports 294 (1998) 1.
\bibitem{wwa}
       C. F.Weizs\"acker, Z. Phys. 88 (1934) 612;
       E. J.Williams, Phys. Rev. 45 (1934) 729.
\bibitem{frixione} S. Frixione, M. L. Mangano, P. Nason, G. Ridolfi,
  Phys. Lett. B319 (1993) 339.
\bibitem{hagedorn}
       R.~Hagedorn, Riv. Nuovo Cim. 6 (1983) 1.
\bibitem{ua1-1}
       UA1 Collab., C. Albajar et al., Nuc. Phys.  B 335 (1990) 261.
\bibitem{cdf} CDF Collab., F. Abe et al., Phys. Rev. Lett. 77 (1996) 438. 
\bibitem{kniehl}
       F. M. Borzumati, B. A. Kniehl and G. Kramer, Z.~Phys. C 59 (1993) 341;\\
       B. A. Kniehl and G. Kramer, Z. Phys. C62 (1994) 53;\\
       B. A. Kniehl, hep-ph/9709261 and Proceedings of the Ringberg Workshop
       ``New Trends in HERA Physics", Ringberg Castle, Germany, 25-30 May 1997.

\bibitem{blobel} V. Blobel, DESY 84-118 and Proceedings of the CERN
School of Computing, Aiguablava (Spain) CERN 1985.


\end{thebibliography}
\end{document}